\documentclass[12pt]{amsart}
\usepackage{amssymb}
\usepackage{verbatim}
\usepackage[usenames]{color}
\usepackage{hyperref}
\usepackage{url}

\newtheorem{thm}{Theorem}%[section]
\newtheorem{prop}[thm]{Proposition}
\newtheorem{la}[thm]{Lemma}

\theoremstyle{definition}
\newtheorem{df}[thm]{Definition}

\theoremstyle{remark}

\newtheorem{rmk}[thm]{Remark}

\newenvironment{ls}{\begin{itemize}}{\end{itemize}}

\newenvironment{pf}{\begin{proof}}{\end{proof}}

\newcommand{\scr}[1]{\ensuremath{\mathcal {#1}}}

\newcommand{\bbb}[1]{\ensuremath{\mathbb {#1}}}

\renewcommand{\phi}{\varphi}

\newcommand{\notarrow}{\kern .42em\not\kern -.42em\longrightarrow}

\newcommand{\ket}[1]{\ensuremath{|#1\rangle}}
\newcommand{\bra}[1]{\ensuremath{\langle#1|}}

\newcommand{\nm}[1]{\Vert #1 \Vert}
\newcommand{\cv}[1]{\ensuremath{\text{Conv}(#1)}}
\newcommand{\af}[1]{\ensuremath{\text{Aff}(#1)}}

\newcommand{\noprint}[1]{\relax}

\title{Spekkens's Symmetric No-Go Theorem}
\author{Andreas Blass}
\address{Mathematics Department\\
University of Michigan\\
Ann Arbor, MI 48109--1043, U.S.A.}
\email{ablass@umich.edu}
\author{Yuri Gurevich}
\address{Microsoft Research\\
One Microsoft Way\\
Redmond, WA 98052, U.S.A.}
\email{gurevich@microsoft.com}

\begin{document}

\maketitle

In \cite{spekkens}, Spekkens clarifies the ways in which classical
theories differ from quantum mechanics.  He improves the traditional
notions of non-negativity of Wigner-style quasi-probability
distributions and non-contextuality of observations.  He argues that
the improvements more accurately capture what a classical universe
would look like.  Thus, both of these improved notions serve to
distinguish quantum theory from classical theories, in particular from
theories that use hidden variables in an attempt to explain the
results of quantum mechanics on a classical basis.  Spekkens then
shows that the two improved notions are equivalent to each other.

Spekkens's improvements of non-negativity and non-contextuality
emphasize the involvement of both preparations and measurements.  In
the second part of \cite{spekkens}, Spekkens provides what he calls an
even-handed approach to a no-go theorem.  The theorem asserts that the
requirement of non-contextuality (or equivalently of non-negativity)
prevents a theory from matching the predictions of quantum mechanics;
in other words, non-contextual hidden-variable theories can't succeed.
``Even-handed'' means that the proof treats preparations and
measurements in a symmetrical way.

The paper \cite{spekkens} contains some minor inaccuracies and one
false claim, in the proof of the no-go theorem.  The false claim is
``that a function $f$ that is convex-linear on a convex set $S$ of
operators that span the space of Hermitian operators (and that takes
value zero on the zero operator if the latter is in $S$) can be
uniquely extended to a linear function on this space.''
Unfortunately, this claim, early in the proof, is used in an essential
way in the rest of the argument.  In this note, we analyze carefully Spekkens's proof of the no-go theorem, explain the inaccuracies, reduce the task of proving the no-go theorem to the special case of a single qubit and then prove the special case. This gives us a complete proof of
Spekkens's no-go theorem. An alternative proof of the no-go theorem is given in the series of papers \cite{Ferry2008,Ferry2009,Ferry2010}.

\section{Definitions}

Spekkens defines  a \emph{quasiprobability representation} of a
quantum system by the following features.
\begin{ls}
\item[QPR1] Every density operator $\rho$ is represented by a
  normalized and real-valued function $\mu_\rho$ on a measurable space
  $\Lambda$.
\item[QPR2] Every positive operator-valued measure (POVM) $\{E_k\}$ is
  represented by a set $\{\xi_{E_k}\}$ of real-valued functions on
  $\Lambda$ that sum to the unit function on $\Lambda$.  (The trivial
  POVM $\{I\}$ is represented by $\xi_I(\lambda)=1$, and the zero
  operator is represented by the zero function.)
\item[QPR3] For all density operators $\rho$ and all POVM elements
  $E_k$, we have $\text{Tr}(\rho E_k)=\int
  d\lambda\,\mu_\rho(\lambda)\xi_{E_k}(\lambda)$.
\end{ls}
A quasiprobability representation is called \emph{nonnegative} if all
the functions $\mu_\rho$ and $\xi_E$ take only nonnegative values.

We begin our analysis by looking carefully at the notions used in this
definition of quasiprobability representation and clarifying some
aspects of the definition.

\subsection{Density operator}
Within the definition of quasiprobability representation, Spekkens
explains ``density operator'' as ``a positive trace-class operator on
a Hilbert space \scr H''.   Although ``trace-class'' implies that the
operator $\rho$ has a well-defined trace, Spekkens presumably intended
more, namely that the trace $\text{Tr}(\rho)$ should be equal to 1.
This would conform with the usual meaning of ``density operator''; it
would also account for the requirement that $\mu_\rho$ be normalized.
If one could multiply $\rho$ by a positive real factor and still have
a density operator, then the associated $\mu_\rho$ should also be
multiplied by the same factor.  From now on, we shall assume that
``trace 1'' is included in the definition of density operator.

It is also worth remembering that ``trace-class'' is important only in
the context of infinite-dimensional Hilbert spaces.  If \scr H is
finite-dimensional, then all (linear) operators on it are in the trace
class.  Spekkens's proof of the no-go theorem does not require an
infinite-dimensional space; it works as long as the dimension of \scr
H is at least 2.  So for many purposes, we need not worry about the
``trace-class'' clause in the definition of density operators.

\subsection{Measurable space}
The phrase ``measurable space'' is standard terminology for a set $X$
together with a $\sigma$-algebra $\Sigma$ of subsets of $X$, the
members of $\Sigma$ being called the measurable sets.  A measurable
space differs from a measure space in that the latter has, in addition
to $X$ and $\Sigma$, a countably additive measure defined on all the
measurable sets.

We believe that Spekkens intends $\Lambda$ to be not merely a
measurable space but a measure space.  He uses the formula
$\int\mu_\rho(\lambda)\,d\lambda=1$ as the definition of the
requirement in QPR1 that $\mu_\rho$ be normalized.  This integral and
the one in clause QPR3 of the definition of quasiprobability
representation both presuppose the presence of a measure to make sense
of $d\lambda$.  They also presuppose that the functions $\mu_\rho$ are
measurable.

An alternative modification to make sense of these integrals would be
to change the requirement that $\rho$ is represented by a
\emph{function} and to require instead that it be represented by a
\emph{measure}, say $\nu_\rho$.  The notation
$\mu_\rho(\lambda)\,d\lambda$ could then be taken to be syntactic
sugar for $d\nu_\rho(\lambda)$.  This alternative approach has, as far
as we can see, two disadvantages and two advantages.  The first
disadvantage is that it requires us to understand Spekkens's notation
$\mu_\rho(\lambda)\,d\lambda$, which looks like a standard notation,
as syntactic sugar for something rather different.  The second is that
it explicitly contradicts Spekkens' assertion that $\mu_\rho$ should
be a function.  The first advantage is that it preserves Spekkens's
convention that $\Lambda$ is merely a measurable space, not a measure
space. The second advantage is that it is more general.  In the
approach with an \emph{a priori} given measure $d\lambda$, multiplying
it by functions $\mu_\rho$ produces only those measures
$\mu_\rho(\lambda)\,d\lambda$ that are absolutely continuous with
respect to $d\lambda$.  The alternative approach allows arbitrary
measures (on the given $\sigma$-algebra $\Sigma$) without any
requirement of absolute continuity.

\subsection{Positive operator-valued measures}
The second defining feature, QPR2, of a quasiprobability
representation represents positive operator valued
measures\footnote{We follow Spekkens's usage of ``POVM'' to refer to a
  discrete set of operators.  This usage agrees with the standard text
  \cite{nc}.  There is a generalization, involving operator-valued
  measures; see for example \cite{wiki-povm}.  For our purposes, the
  simpler version is adequate, since the no-go theorem for these
  simpler POVMs implies the theorem for the broader class.}  $\{E_k\}$
by sets of functions $\xi_{E_k}$.  The elements $E_k$ of a POVM are
positive Hermitian operators such that $I-E_k$ is also positive.  That
is, the spectrum of $E_k$ lies in the interval $[0,1]$ of the real
line.  Conversely, any such operator occurs as a member of some POVM,
and usually as a member of many POVMs.  Specifically, if $E$ is a
positive Hermitian operator and $I-E $ is also positive, then
$\{E,I-E\}$ is a POVM; unless $E=I$, we can replace $I-E$ in this POVM
by two or more positive operators whose sum is $I-E$, thereby
obtaining other POVMs containing $E$.

The question arises whether the function $\xi_{E_k}$ in a
quasiprobability representation can depend on the POVM from which
$E_k$ was taken or must depend only on the operator $E_k$ itself.  The
wording of the definition suggests the former, while the notation
$\xi_{E_k}$ suggests the latter.  Fortunately for our purposes,
Spekkens's definition of ``measurement noncontextuality'' requires that
$\xi_{E_k}$ ``depends only on the associated POVM \emph{element}
$E_k$'' (italics added).  Since our goal in this paper, the no-go
theorem, is about noncontextual representations, we can safely follow
the notation and assume that $\xi_E$ depends only on $E$, not on the
POVM in which it occurs (and, a fortiori, not on the measurement
process by which that POVM is realized).

\section{An additional hypothesis}

At the beginning of his proof of the no-go theorem, Spekkens notes
that a mixture $\rho=\sum_jw_j\rho_j$ of density operators $\rho_j$
with weights $w_j$ can be prepared by first randomly choosing one
value of $j$ from the probability distribution $\{w_j\}$ and then
preparing $\rho_j$.  He infers that ``clearly''
$\mu_\rho(\lambda)=\sum_jw_j\mu_{\rho_j}(\lambda)$.

Although this inference is highly plausible and natural on physical
grounds, it does not follow from just the definition of
quasiprobability distribution as quoted above.  Suppose that the
functions $\xi_E$ do not span the whole space of square-integrable
functions on $\Lambda$, so that there is a function $\sigma$
orthogonal to all of these $\xi_E$'s, where ``orthogonal'' means that
$\int\sigma(\lambda)\xi_E(\lambda)\,d\lambda=0$.  One could modify the
$\mu_\rho$ functions by adding to each one some multiple of $\sigma$,
obtaining $\mu'_\rho=\mu_\rho+c_\rho\sigma$ and still satisfying the
definition of quasiprobability representation.  Here the coefficients
$c_\rho$ can be chosen arbitrarily for each density operator $\rho$.
By choosing them in a sufficiently incoherent way, one could arrange
that $\mu'_\rho(\lambda)\neq\sum_jw_j\mu'_{\rho_j}(\lambda)$.

If, on the other hand, the $\xi_E$'s do span the whole space of
functions on $\Lambda$, then Spekkens's desired equation
$\mu_\rho(\lambda)=\sum_jw_j\mu_{\rho_j}(\lambda)$ does follow, for
all but a measure-zero set of $\lambda$'s, because the two sides of
the equation must give the same result when integrated against any
$\xi_E$.

Unfortunately, nothing in the definition of quasiprobability
representations requires the $\xi_E$'s to span the whole space.  For
example, given any quasiprobability representation, we can obtain
another, physically equivalent one as follows.  Replace $\Lambda$ by
the disjoint union $\Lambda_1\sqcup\Lambda_2$ of two copies of
$\Lambda$.  Define the measure of any subset of
$\Lambda_1\sqcup\Lambda_2$ to be the average of the original measures
of its intersections with the two copies of $\Lambda$.  Define all the
functions $\mu_\rho$ and $\xi_E$ on the new space by simply copying
the original values on both of the $\Lambda_i$'s.  The result is a
quasiprobability representation in which the $\xi_E$'s span only the
space of functions that are the same on the two copies of $\Lambda$.

The result of this discussion is that, in order to prove the no-go
theorem along the lines proposed by Spekkens, we must add an
additional hypothesis about mixtures of densities.  There is a similar
assumption for mixtures of measurements.

\smallskip\noindent\textbf{Convex-linearity Hypothesis:} Let $\{w_j\}$ be a
probability distribution on a set of indices $j$.
\begin{ls}
\item If $\rho=\sum_jw_j\rho_j$, then
  $\mu_\rho(\lambda)=\sum_jw_j\mu_{\rho_j}(\lambda)$.
\item If $E=\sum_jw_jE_j$, then
  $\xi_E(\lambda)=\sum_jw_j\xi_{E_j}(\lambda)$.
\end{ls}

\smallskip
\noindent
This hypothesis is exactly statements (7) and (8) in \cite{spekkens}.
The name of the hypothesis refers to the following terminology, which
we shall need again later.

\begin{df}
Let $C$ be a convex set in a real vector space $V$, and let $f$ be a
function from $C$ into another real vector space $W$.  Then $f$ is
\emph{convex-linear} on a subset $S$ of $C$ if
\[
f(a_1v_1+\cdots+a_nv_n)=a_1f(v_1)+\cdots+a_nf(v_n)
\]
for all vectors $v_1,\dots,v_n\in S$ and all nonnegative numbers
$a_1,\dots,a_n$ with $a_1+\cdots+a_n=1$.
\end{df}

Thus, the convex-linearity hypothesis says that the functions
$\rho\mapsto\mu_\rho$ and $E\mapsto\xi_E$ are convex-linear on the
sets of density matrices and POVM elements, respectively.

\section{The no-go theorem}

On an intuitive level, the no-go theorem asserts that nonnegative
quasiprobability representations\footnote{These are equivalent to
  noncontextual ontological models, as Spekkens shows in the earlier
  sections of \cite{spekkens}.} subject to the convex-linearity
hypothesis cannot reproduce the predictions of quantum mechanics.  A
considerable amount of agreement with quantum mechanics is already
built into the definition of quasiprobability representations.
Specifically, the equation $\text{Tr}(\rho
E_k)=\int\mu_\rho(\lambda)\xi_{E_k}(\lambda)\,d\lambda$ says that the
expectation of $E_k$ in state $\rho$ is the same whether computed by
the quantum formula $\text{Tr}(\rho E_k)$ or as an average using the
functions $\mu_\rho$ and $\xi_{E_k}$ from the quasiprobability
representation.  Spekkens's no-go theorem asserts that there is no
nonnegative quasiprobability representation satisfying
convex-linearity.

A small technical point is that the no-go theorem presupposes that the
quantum mechanics is non-trivial.  Quantum mechanics on Hilbert spaces
of dimensions 0 or 1 is classical (and trivial), so we must assume
that we are dealing with a Hilbert space \scr H of dimension at least
2.  An inspection of Spekkens's argument reveals that he never uses
any stronger assumptions about \scr H.  Thus, the no-go theorem can be
formally stated as follows.

\begin{thm}     \label{no-go}
  For a Hilbert space \scr H of dimension at least two, there is no
  way to define nonnegative $\mu_\rho$, for all density operators
  $\rho$, and to define nonnegative $\xi_E$, for all positive
  Hermitian operators $E$ with $I-E$ positive, so as to satisfy both
  the definition of a quasiprobability representation and the
  convex-linearity hypothesis.
\end{thm}

\section{Reduction to two dimensions}

In this section, we reduce the task of proving Spekkens's no-go
theorem to the special case where \scr H has dimension 2.  (In the
terminology of quantum computing, \scr H represents a single qubit.)
More generally, we show that, if there were a nonnegative
quasiprobability representation satisfying convex-linearity
for some Hilbert space \scr H, then there would also be such a
representation, using the same measure space $\Lambda$, for any
nonzero, closed subspace $\scr H'$ of \scr H.

To see this, suppose functions $\mu_\rho$  (for all $\rho$) and
$\xi_E$ (for all $E$) constitute such a representation for \scr H.
Let $i:\scr H'\to\scr H$ be the inclusion map (the identity map of
$\scr H'$ regarded as a map into \scr H), and let $p:\scr H\to\scr H'$
be the orthogonal projection map (sending each vector in $\scr H'$ to
itself and sending each vector orthogonal to $\scr H'$ to 0).  Also,
fix some unit vector $\ket\alpha\in\scr H'$.

Each density operator $\rho$ on $\scr H'$ gives rise to a density
operator $\bar\rho=i\circ\rho\circ p$ on \scr H.  For pure states,
this amounts to just considering a state vector in $\scr H'$ as a
vector in the larger Hilbert space \scr H.  For mixed states, the
extension preserves averages.  We begin defining a quasiprobability
representation for $\scr H'$ by setting $\mu'_\rho=\mu_{\bar\rho}$.
We note that this is a normalized nonnegative real-valued function on
$\Lambda$, and that it satisfies the part of convex-linearity
that refers to the representations of densities.

It is tempting to proceed exactly analogously with POVM elements $E$
and their representing functions $\xi_E$.  That procedure doesn't
quite work, because the definition of quasiprobability representation
imposes a specific requirement on $\xi_I$, where $I$ is the identity
operator.  Unfortunately, if $I$ is the identity operator on $\scr
H'$, then $i\circ I\circ p$ is not the identity operator on \scr H.
So we must proceed slightly differently, and it is here that the fixed
unit vector $\ket\alpha$ will be useful.

Given a POVM element $E$ on $\scr H'$, i.e., a positive, Hermitian
operator such that $I-E$ is also positive, we define $\bar E$ to be
the unique linear operator on \scr H such that
\[
\bar E\ket\psi =
\begin{cases}
E\ket\psi & \text{if }\psi\in\scr H',\\
\bra\alpha E\ket\alpha\ket\psi & \text{if }\ket\psi\bot\scr H'.
\end{cases}
\]
In other words, $\bar E$ agrees with $E$ on $\scr H'$ and with a
scalar multiple of the identity on the orthogonal complement of $\scr H'$,
the multiplier of the identity being $\bra\alpha E\ket\alpha$.    This
extension process produces POVM elements for \scr H; indeed, if a set
$\{E_k\}$ of operators is a POVM for $\scr H'$, then $\{\bar E_k\}$ is
a POVM for \scr H.  Furthermore, the extension process sends the
identity and zero operators on $\scr H'$ to the identity and zero
operators on \scr H, and the process respects weighted averages.

We continue the definition of a quasiprobability representation for
$\scr H'$ by setting $\xi'_E=\xi_{\bar E}$ for all POVM elements $E$
on $\scr H'$.  The remarks above immediately imply that these
functions $\xi'_E$ are as required by the second part, QPR2, of the
definition of quasiprobability representation, that they are
nonnegative, and that they satisfy the relevant part of the
convex-linearity hypothesis.

To verify the last part, QPR3, of the definition of quasiprobability
representation, we observe that, for any density operator $\rho$ and
POVM element $E$ on $\scr H'$, the extensions $\bar\rho$ and $\bar E$
agree with $\rho$ and $E$ on $\scr H'$, while on the orthogonal
complement of $\scr H'$, $\bar\rho$ acts as zero and $\bar E$ acts as
a scalar multiple of the identity.  It follows immediately that
$\text{Tr}(\bar\rho\bar E)=\text{Tr}(\rho E)$, and therefore
\[
\text{Tr}(\rho E)=\int d\lambda\,\mu'_\rho(\lambda)\xi'_E(\lambda),
\]
as required.

This completes the proof that nonnegative quasiprobability
representations subject to convex-linearity can be
``restricted'' to nonzero, closed subspaces of the original Hilbert
space.  Therefore, it suffices to prove the no-go theorem in the
special case where \scr H has dimension 2.

\begin{rmk}
  By concentrating on the case of dimension 2, we gain two advantages.
  First, we can avoid some technicalities that would arise for
  infinite-dimensional Hilbert spaces.  Second, we obtain a more
  concrete picture of the relevant spaces of density operators and
  measurements.  (The first of these advantages would result from
  reduction to any finite number of dimensions; the second benefits
  specifically from dimension 2.)
\end{rmk}

\section{Convex-linear transformations}
Spekkens asserts that, if a function $f$ is convex-linear on a convex
set \scr S of operators that span the space of Hermitian operators
(and $f$ takes the value zero on the zero operator if the latter is in
\scr S), then $f$ can be uniquely extended to a linear function on
this space.  Unfortunately, such a linear extension need not exist in
the general case, when zero is not in \scr S.\footnote{Spekkens gives
  a formula purporting to define a linear extension of $f$ in general,
  but it is not well-defined because it involves some arbitrary
  choices.  He also gives, in footnote~18 of the newer version
  \cite{spekkens2} of his paper, an argument purporting to show that
  his formula is independent of those choices, but that argument
  fails.  It involves dividing by an appropriate constant $C$ to turn
  two nonnegative linear combinations, the two sides of an equation,
  into convex combinations so that the assumption of convex-linearity
  can be applied.  But the necessary divisor $C$ may need to be
  different for the two sides of the equation.}  For a simple example,
consider the function that is identically 1 on an \scr S that spans
the space of Hermitian operators, does not contain 0, but does contain
two orthogonal projections and their sum.

The correct version of the result extends $f$ not to a linear function but to translated-linear function, i.e., a composition of translations and a linear function.  The rest of this section is devoted to a proof of this fact, in somewhat greater generality than we need.  It applies to arbitrary real vector spaces; that the space consists of Hermitian operators is irrelevant.

The \emph{convex hull}, \cv S, of a subset $S$ of a real vector space
$V$ consists of the convex combinations $a_1v_1+\cdots+a_nv_n$ of
vectors $v_1,\dots,v_n\in S$ where $a_1+\cdots+a_n=1$ and every
$a_i\geq0$.  The \emph{affine hull}, \af S, of $S$ consists of the
affine combinations $a_1v_1+\cdots+a_nv_n$ of vectors
$v_1,\dots,v_n\in S$ where $a_1+\cdots+a_n=1$ but some coefficients
$a_i$ may be negative.

A set is \emph{convex} if it contains all the convex combinations of
its members; similarly, it is an \emph{affine} space if it contains
all the affine combinations of its members.  An easy computation shows
that convex hulls are convex and affine hulls are affine spaces; that
is $\cv{\cv S}=\cv S$ and $\af{\af S}=\af S$.

An affine space $A$ in a vector space $V$ is said to be \emph{parallel}
to a linear subspace $L$ of $V$ if $A=u_0+L=\{u_0+v:v\in L\}$ for some
$u_0\in V$.
It is easy to see that, if an affine space $A$ is
parallel to a linear space $L$ as above, then (i)~$L$ is unique,
(ii)~$u_0\in A$, (iii)~any vector in $A$ can play the role of the
translator $u_0$, and (iv)~$A$ is either equal to $L$ or disjoint from
$L$.

\begin{la}[\S1 in \cite{Rockafellar}]\label{Rockafellar}
Any affine subspace $A$ of a real vector space $V$ is parallel to a
linear subspace $L$ of $V$.
\end{la}

In other words, any affine subspace is a translation of a linear
subspace.  For example, in $\bbb R^2$, we have that
$\mathrm{Aff}\{(0,1),(1,0)\}$ is parallel to the diagonal $y=-x$, and
$\mathrm{Aff}\{(0,1),(1,0),(1,1)\}$ is (and thus is parallel to) $\bbb
R^2$.

\begin{proof}
If $A$ contains the zero vector $\vec0$ then it is a linear
subspace. Indeed, if $v\in A$ then any multiple $av = av +
(1-a)\vec0 \in A$. And if $u,v\in A$ then $u+v = 2(\frac12u +
\frac12v)\in A$.

For the general case, let $u_0$ be any vector in the affine space $A$.
It suffices to show that $L = \{v-u_0: v\in A\}$ is an affine space,
because then the preceding paragraph shows that it is a linear space,
and clearly $A=u_0+L$. Any affine combination
$a_1(v_1-u_0)+\cdots+a_n(v_n-u_0)$ of vectors in $L$ (so the $v_i$ are
in $A$ and the sum of the $a_i$ is 1) can be rewritten as
$(a_1v_1+\cdots+a_nv_n)-u_0$, which is in $L$.
\end{proof}

Let $V$ and $W$ be real vector spaces, $S$ a subset of $V$, $C=\cv S$
its convex hull, and $A=\af S$ its affine hull.  Recall that a
transformation $f:C\to W$ is convex-linear on $S$ if
\[
  f(a_1v_1 +\cdots+ a_nv_n) = a_1f(v_1) +\cdots+ a_nf(v_n)
\]
for any convex combination $a_1v_1 +\cdots+ a_nv_n$ of vectors $v_i$
from $S$.  A transformation $f:A\to W$ is \emph{translated-linear} if
it has the form $f(v)=w_0+h(v-u_0)$ for some $w_0\in W$, some $u_0\in
A$, and some linear function $h:L\to W$ defined on the linear space
$L=A-u_0$ parallel to $A$.

\begin{prop} \label{translinear} With notation as above, any
  transformation $f:C\to W$ that is convex-linear on $S$ has a unique
  extension to a translated-linear function on $A$.
\end{prop}

\begin{pf}
Notice first that translations $v\mapsto v-u_0$ and linear functions both
preserve affine combinations.  A translated-linear function, being the
composition of two translations and a linear function, therefore also
preserves affine combinations.  This observation implies the
uniqueness part of the proposition.  Indeed, every element of $A$ is
an affine combination $a_1s_1+\cdots+a_ns_n$ of elements of $S$, and
therefore any translated-linear extension of $f$ must map it to
$a_1f(s_1)+\cdots+a_nf(s_n)$.

To prove the existence part of the proposition, it will be useful to
work with the graphs of functions.  For any function $g:S\to W$ with
$S\subseteq V$, its \emph{graph} is the subset of $V\oplus W$
consisting of the pairs $(s,g(s))$ for $s\in S$.\footnote{In
  set-theoretic foundations, a function is usually defined as a set of
  ordered pairs, and so $g$ is the same thing as its graph.}  We
record for future reference that the graph of $g$ is a linear subspace
of $V\oplus W$ if and only if the domain of $g$ is a linear subspace
of $V$ and $g$ is a linear transformation from that domain to $W$.  We
also note that the projection $\pi:V\oplus W\to V:(v,w)\mapsto v$ is a
linear transformation that sends the graph of any $g$ to the domain of
$g$.

In the situation of the proposition, let $f:C\to W$ be a
transformation that is convex-linear on $S$, and let $F\subseteq
V\oplus W$ be its graph.  Also, let $F^-$ be the graph of the
restriction of $f$ to $S$.  Notice that the convex-linearity of $f$ on
$S$ means exactly that $F$ is the convex hull of $F^-$.  It follows
that $F$ and $F^-$ have the same affine hull, because
\[
\af F=\af{\cv{F^-}}\subseteq\af{\af{F^-}}=\af{F^-}\subseteq\af F.
\]
We claim that this affine hull $\af{F^-}$ is the graph of a function;
that is, it does not contain two distinct elements $(v,w)$ and
$(v,w')$ with the same first component $v$.  To see this, suppose we
had two such elements in $\af F=\af{F^-}$, say
\[
(v,w)=a_1(s_1,f(s_1))+\cdots+a_m(s_m,f(s_m))
\]
and
\[
(v,w')=b_1(t_1,f(t_1))+\cdots+b_n(t_n,f(t_n)),
\]
where all the $s_i$'s and $t_j$'s are in $S$ and where
\begin{equation}\label{rep1}
     a_1 +\cdots+ a_m = b_1 +\cdots+ b_n,
\end{equation}
because both sides are equal to 1. So we have
\begin{equation}\label{rep2}
    a_1s_1 +\cdots+ a_ms_m = b_1t_1 +\cdots+ b_nt_n ,
\end{equation}
because both sides are equal to $v$, and we want to prove $w=w'$,
i.e.,
\begin{equation}\label{rep3}
    a_1f(s_1) +\cdots+ a_mf(s_m) = b_1f(t_1) +\cdots+ b_nf(t_n ).
\end{equation}
In the special case where all coefficients $a_i$ and $b_j$ are $\ge0$,
vector $v$ is in $C$ and both sides of \eqref{rep3} are equal to
$f(v)$. The general case reduces to this special case as follows. In
all three equations \eqref{rep1}--\eqref{rep3}, move every summand
with a negative coefficient to the other side, and then divide the
resulting equations by the left part of the rearranged equation
\eqref{rep1}. As a result we return to the special case already
treated. Since the old version of \eqref{rep3} follows from the new
one, this completes the proof of our claim that $\af F=\af{F^-}$ is
the graph of a function.

By Lemma~\ref{Rockafellar}, the affine space $\af F$ is parallel to a
linear subspace $H$ of $V\oplus W$, say $\af F=(u_0,w_0)+H$, where
$u_0\in V$ and $w_0\in W$.  From the fact that \af F is the graph of a
function, it follows immediately that $H$ is also the graph of a
function.  Indeed, if $H$ contains $(v,w)$ and $(v,w')$, then \af F
contains $(v-u_0,w-w_0)$ and $(v-u_0,w'-w_0)$, so $w-w_0=w'-w_0$ and
$w=w'$.

Let $h$ be the function whose graph is $H$.  Because $H$ is a linear
subspace of $V\oplus W$, we know that $h$ is a linear transformation
from some linear subspace $L$ of $V$ into $W$.

The fact that $(u_0,w_0)+H=\af F$ tells us, by applying the linear projection
$\pi:V\oplus W\to V$, that $u_0+L$ equals
\[
\pi(\af F)=\af{\pi(F)}=\af C=A,
\]
where the first equality comes from linearity of $\pi$ and the second
from the fact that $F$ is the graph of the function $f$ whose domain
is $C$.  So $A$ is parallel to the linear subspace $L$ of $V$.
Furthermore, for each $v\in C$, we have
\[
(v,f(v))\in F\subseteq \af F=(u_0,w_0)+H,
\]
so $(v-u_0,f(v)-w_0)$ is in the graph $H$ of $h$.  That is,
$h(v-u_0)=f(v)-w_0$ and so $f(v)=w_0+h(v-u_0)$.  Thus, the
translated-linear function $v\mapsto w_0+h(v-u_0)$ is the desired
extension of $f$.
\end{pf}

\begin{rmk}
A linear function $h$ on a subspace $L$ of a vector space $V$ can be
extended to a linear function $\bar h$ on all of $V$.  Extend any
basis of $L$ to a basis of $V$, define $\bar h$ arbitrarily on the
new basis vectors that are not in $L$, and extend the resulting
function by linearity to all of $V$.

For transformations defined on
all of $V$, we have a simpler formula for translated-linear functions,
because
\[
w_0+\bar h(v-u_0)=w_0+\bar h(v)-\bar h(u_0)=\bar h(v)+w_1,
\]
where $w_1=w_0-\bar h(u_0)$.

On the other hand, in contrast to Proposition~\ref{translinear}, this
$\bar h$ is not unique (unless $L=V$).

Also, in the case of infinite-dimensional spaces, the extension
process requires the axiom of choice (to extend bases) and need not be
well-behaved with respect to natural topologies on the vector spaces.
\end{rmk}

\section{Density operators and POVM elements in two dimensions}

In this section, we recall the form of density operators and POVM
elements in the case where \scr H is two-dimensional.  In this case, a
basis for the Hermitian operators on \scr H is given by the identity
and the three Pauli matrices
\[
I=
\begin{pmatrix}
  1&0\\0&1
\end{pmatrix},\quad X=
\begin{pmatrix}
  0&1\\1&0
\end{pmatrix},\quad Y=
\begin{pmatrix}
  0&-i\\i&0
\end{pmatrix},\quad Z=
\begin{pmatrix}
  1&0\\0&-1
\end{pmatrix}.
\]
It will be convenient to use vector notation, denoting the triple  of
matrices $(X,Y,Z)$ by $\vec X$.  Then the general Hermitian matrix
looks like
\[
wI+xX+yY+zZ=wI+\vec x\cdot\vec X,
\]
where $w$ and the three components of $\vec x$ are real numbers.  The
eigenvalues of this Hermitian matrix are
\[
w\pm\sqrt{x^2+y^2+z^2}=w\pm\nm{\vec x}
\]
In particular, the trace of this matrix is $2w$, and the matrix is
positive if and only if $w\geq\nm{\vec x}$.

Density matrices are the Hermitian, positive matrices of trace 1, so
they have the form
\[
\rho=\rho(\vec x)=\frac12(I+\vec x\cdot\vec X),
\]
where $\nm{\vec x}\leq 1$.  As indicated by the notation, we
parametrize these density matrices by three-component vectors $\vec x$
of norm $\leq 1$.  The three-dimensional ball that serves as the
parameter space here is called the Bloch sphere (with its interior).

Similarly, POVM elements have the form
\[
E=E(m,\vec p)=mI+pX+qY+rZ=mI+\vec p\cdot\vec X
\]
with
\[
\nm{\vec p}\leq m\leq 1-\nm{\vec p}
\]
(because $E$ and $I-E$ are positive operators) and therefore $\nm{\vec
  p}\leq\frac12$.    The parameter space here,
consisting of all four-component vectors satisfying these
inequalities, is a double cone over a three-dimensional ball of radius
$\frac12$.

We record for future reference the traces
\[
\text{Tr}(I)=2,\quad\text{Tr}(X)=\text{Tr}(Y)=\text{Tr}(Z)=0
\]
and the multiplication table
\[
XY=-YX=iZ, \quad YZ=-ZY=iX,\quad ZX=-XZ=iY,
\]
and
\[
 X^2=Y^2=Z^2=I.
\]
From these facts, it is easy to compute that
\[
\text{Tr}(\rho(\vec x)E(m,\vec p))=m+\vec x\cdot\vec p,
\]
where the factor $\frac12$ in the definition of $\rho(\vec x)$ has
cancelled the factor 2 arising from $\text{Tr}(I)$.

\section{Quasiprobability representation}

Finally, we are ready to prove Theorem~\ref{no-go}.

Suppose, toward a contradiction, that we have a nonnegative
quasiprobability representation satisfying convex-linearity, for a
two-dimensional \scr H.  In view of Proposition~\ref{translinear}, we
know that
\[
\mu_{\rho(\vec x)}(\lambda)=\vec x\cdot\vec A(\lambda)+C(\lambda)
\]
and
\[
\xi_{E(m,\vec p)}= \vec p\cdot\vec B(\lambda)+m D(\lambda)+F(\lambda)
\]
for some nine functions $A_i(\lambda), B_i(\lambda), C(\lambda),
D(\lambda), F(\lambda)$ where the index $i$ ranges from 1 to 3.  (The
``translated'' part of ``translated-linear'' accounts for $C$ and
$F$.)

The definition of quasiprobability representation leads to some
simplifications.  $E(0,\vec0)$ is the zero operator, whose
associated $\xi$ function is required to be identically zero.  That
gives us $F(\lambda)=0$ for all $\lambda$, so we can simply omit $F$
from the formula for $\xi$.

Also, $E(1,\vec 0)$ is the identity operator, whose associated $\xi$
function is required to be identically 1.  That gives us
$D(\lambda)=1$ for all $\lambda$.  So we can simplify the $\xi$
formula above to read
\[
\xi_{E(m,\vec p)}= \vec p\cdot\vec B(\lambda)+m.
\]

Next, consider the requirement that
\[
\text{Tr}(\rho(\vec x)E(m,\vec p))=\int\xi_{E(m,\vec
  p)}\mu_{\rho(\vec x)}\,d\lambda.
\]
We already evaluated the trace on the left side of this equation at
the end of the preceding section.  The integral on the right side is
\[
\int[(\vec p\cdot\vec B(\lambda))(\vec x\cdot\vec A(\lambda))+
(\vec p\cdot\vec B(\lambda))C(\lambda)+
m(\vec x\cdot\vec A(\lambda))+mC(\lambda)]\,d\lambda.
\]
Comparing the trace and the integral, and equating coefficients of the
various monomials in $m$, $\vec p$, and $\vec x$, we find that
\begin{align}
\int B_i(\lambda)A_j(\lambda)\,d\lambda&=\delta_{i,j}, \label{2}\\
\int B_i(\lambda)C(\lambda)\,d\lambda&=0, \label{3}\\
\int A_i(\lambda)\,d\lambda&=0, \text{and} \label{4}\\
\int C(\lambda)\,d\lambda&=1.\label{5}
\end{align}

Next, we extract as much information as we can from the assumption
that all the functions $\mu_\rho$ and $\xi_E$ are nonnegative.

In the case of $\xi_E$, this means that, as long as $\nm{\vec p}\leq
m,1-m$ (so that $E(m,\vec p)$ is a POVM element), we must have $m+\vec
p\cdot\vec B(\lambda)\geq 0$ for all $\lambda$.  Temporarily consider
a fixed $\lambda$ and a fixed $m\in[0,\frac12]$. To get the most
information out of the inequality $m+\vec p\cdot\vec B(\lambda)\geq
0$, we choose the ``worst'' vector $\vec p$, i.e., we make $\vec
p\cdot\vec B(\lambda)$ as negative as possible, by choosing $\vec p$
in the opposite direction to $\vec B(\lambda)$ and with the largest
permitted magnitude, namely $m$.  That is, we take
\[
\vec p=-\frac m{\nm{\vec B(\lambda)}}\vec B(\lambda)
\]
so that our inequality becomes $0\leq m(1-\nm{\vec B(\lambda)})$, and
therefore
\[
\nm{\vec B(\lambda)}\leq1
\qquad\text{for all }\lambda.
\]
Repeating the exercise for $m\in[\frac12,1]$ gives no new information.

So we turn to the case of $\mu_{\rho(\vec x)}$, for which the
nonnegativity requirement reads
\[
\vec x\cdot\vec A(\lambda)+C(\lambda)\geq 0.
\]
For each fixed $\lambda$, we consider the ``worst" $\vec x$, namely a
vector $\vec x$ in the direction opposite to $\vec A(\lambda)$ and
with the maximum allowed magnitude, namely 1.  So we take
\[
\vec x=-\frac{\vec A(\lambda)}{\nm{\vec A(\lambda)}}
\]
and obtain the inequality $0\leq -\nm{\vec A(\lambda}+C(\lambda)$.
Thus, we have
\[
\nm{\vec A(\lambda)}\leq C(\lambda)\qquad\text{for all }\lambda.
\]
In particular, $C(\lambda)$ is everywhere nonnegative.

A trivial consequence of $\nm{\vec A(\lambda)}\leq C(\lambda)$ is that
$|A_1(\lambda)| \leq C(\lambda)$.  Similarly, a trivial consequence of
$\nm{\vec B(\lambda)}\leq1$ is $|B_1(\lambda)|\leq 1$.  Putting this
information into the $i=j=1$ case of equation~\eqref{2}, and also
using \eqref{5}, we find that
\[
1=\left|\int B_1(\lambda)A_1(\lambda)\,d\lambda\right|
\leq\int |B_1(\lambda)|\cdot|A_1(\lambda)|\,d\lambda
\leq\int 1\cdot C(\lambda)\,d\lambda=1.
\]
So both of the inequalities here must be equalities.  In particular,
$|B_1(\lambda)|=1$ for almost all $\lambda$ except where
$C(\lambda)=0$.

Similarly, we get that, for almost all $\lambda$ except where
$C(\lambda)=0$, we also have $|B_2(\lambda)|=|B_3(\lambda)|=1$ and
therefore $\nm{\vec B(\lambda)}=\sqrt3$.  Since we also know $\nm{\vec
  B(\lambda)}\leq 1$, we must conclude that $C(\lambda)=0$ almost
everywhere.  But that contradicts equation~\eqref{5}, and so the proof
of the no-go theorem is complete.

\end{document}